\documentclass[aps,twocolumn,showpacs]{revtex4}
\usepackage[dvips]{graphics,graphicx,epsfig}
\usepackage{amsfonts}

\begin{document}

\def\beqn{\begin{equation}}
\def\eeqn{\end{equation}}
\def\beqnar{\begin{eqnarray}}
\def\eeqnar{\end{eqnarray}}
\newcommand{\ket}[1]{$\vert${#1}$\rangle$}
\newcommand{\mket}[1]{\vert{#1}\rangle}
\newcommand{\mbra}[1]{\langle{#1}\vert}
\newcommand{\tfrac}[2]{{\textstyle\frac{#1}{#2}}}
\newcommand{\ignore}[1]{}
\def\up{|\uparrow\,\rangle}
\def\dn{|\downarrow\,\rangle}
\def\upd{\langle\, \uparrow |}
\def\dnd{\langle\, \downarrow |}
\newcommand{\mb}[1]{\mbox{\boldmath{$#1$}}}
\def\ba{\begin{array}}
\def\ea{\end{array}}
\newcommand{\wdg}{\! \wedge \!}
\newcommand{\crs}{\! \times \!}
\newcommand{\scp}{\! \ast \!}
\newcommand{\dt}{\! \cdot \!}
\newcommand{\etal}{{\em et al. }}
\newcommand{\eqn}[1]{(\ref{#1})}

\title{Simulation of the Burgers equation by NMR quantum information 
processing}

\author{ Zhiying Chen${}^{1}$,
          Jeffrey Yepez${}^{2}$,
          David G. Cory${}^{1}$\thanks{To whom correspondence should 
be addressed.
          E-mail: dcory@mit.edu } }

\address{ ${}^1$  Department of Nuclear Engineering,
           Massachusetts Institute of Technology, Cambridge, MA  02139 \\
           ${}^2$ Air Force Research Laboratory, Hanscom Field, MA  01731}

\date{April, 2004}

\begin{abstract}
We report on the implementation of Burgers equation as a type-II 
quantum computation on an NMR quantum information processor.  Since 
the flow field evolving under the Burgers equation develops sharp 
features over time, this is a better test of liquid state NMR 
implementations of type-II quantum computers than the previous 
examples using the diffusion equation.  In particular, we show that 
Fourier approximations used in the encoding step are not the dominant 
error.  Small systematic errors in  the
collision operator accumulate and swamp all other errors.  We 
propose, and demonstrate, that the accumulation of this error can be 
avoided to a large extent by  replacing the single collision operator 
with a set
of operators with random errors and similar fidelities. 
Experiments have been implemented on 16 two-qubit
sites for eight successive time steps for the Burgers equation.  
\end{abstract}

\maketitle

It has been suggested that some classical computational problems can 
be solved by using a hybrid classical
quantum device, a type II quantum computer 
\cite{yepez-ijtp98,yepez-ijmpc00a-short}.  Such a device is
essentially an array of small quantum imformation processors (QIP) 
sharing information through classical channels.  NMR has proven to be 
a useful testbed for QIP, and in particular we have shown that
a lattice of parallel QIPs can be mapped onto a spin system by
creating a correspondence between the lattice sites and spatially 
distinct spin ensembles.  A first proof-of-concept for
numerically predicting the time-dependent solution of classical 
partial differential equation with dissipative terms using our NMR technique
was demonstrated for the diffusion equation 
\cite{yepez-pravia-cpc2001,yepez-pravia-pre2002}.

One of the most important challenges to implementing a useful 
type-II quantum architecture is to avoid the accumulation of 
systematic errors.  In the NMR implementations to date there are two 
important sources of systematic errors: (1) a linear approximation 
relating the excited magnetization to the Fourier components of the 
shaped RF pulse; and (2) errors from the repeated collision 
operators.  Here we explore the impact of these errors on a simple 
computation and illustrate a simple means of reducing the accumulated 
error.

The ensemble nature of the spin system allows us to split the sample 
into a spatial array of lattice sites.  Well developed methods from 
magnetic resonance imaging (MRI) allow us to selectively address the 
spins in each of these sites.  Typically the addressing is carried 
out in a space reciprocal to the spatial mapping, called $k$-space, 
where $k$ is the wave-number of the corresponding Fourier components.
The $k$-space formalism \cite{Aaron-kspace} provides a recipe for 
writing a spatially varying spin rotation across an ensemble of spins 
that have been distinguished from each other by a magnetic field 
gradient. The $k$-space formalism is essentially the application of 
shaped radio frequency (RF) pulses in the presence of a linear 
magnetic gradient field as a means of exciting selective frequencies.  For 
most studies the full k-space formalism is not employed and a linear 
approximation is invoked. If the rotation angle of the shaped pulse
is small, then the excited magnetization may be accurately calculated 
only to first order in that angle, and
the excited magnetization is related to the RF waveform simply by a 
Fourier transform.  As a result,
the required RF waveform can also be determined by taking the inverse 
Fourier transform of the desired initial magnetization.
This technique allows us to encode arbitrary magnetization profiles 
spanning the various spatial locations in our
experiment and thereby approximating any desired initial conditions. 
In the previously implemented
diffusion equation, higher order Fourier components of the number 
density are attenuated by the dynamics and the solution is
stable even in the presence of substantial accumulated errors.

To push the development of type-II implementations  we have chosen to 
explore the nonlinear Burgers equation to test the breakdown for the 
Fourier approximation.  Over time, a  shock
front forms and high spatial frequencies in the magnetization profile become 
important and it is these high spatial frequencies that we expect to 
be most sensitive to errors.     The numerical treatment of
the QLG algorithm for the Burgers equation 
\cite{yepez-jstatphy01-short} therefore offers a stronger proof of 
our NMR quantum
computing approach since the effect of the nonlinear convective term 
in the equation generates a sharp edge as a shock develops
in time that is not mimicked by spin relaxation, random 
self-diffusion, nor RF inhomogeneities.
In addition, we demonstrate shock-formation driven by a tunable 
viscosity parameter to show that the width of the shock front is
not determined by implementation imperfections.

The first-order accurate Fourier approximation was expected to be the 
dominant error source in the NMR implementation.
However, NMR simulations with controlled errors shows that the 
systematic error induced by the experimental implementation
of the unitary collision operator associated with the quantum lattice 
gas (QLG) algorithm is the major challenge.
Replacing the single collision operator with a set of operators to randomize errors allows us to improve the robustness of
the implementation.

A quantum lattice gas is a system of quantum particles moving and 
colliding on a discrete spacetime lattice.  This quantum particle
system is isomorphic to a lattice-based qubit system.  The mapping is 
as follows: the probability of a particle residing at
a particular lattice node is equated to the moduli squared of the 
probability amplitude of a qubit at a unique
location being in its excited quantum state.    That is, each spatial 
location that a particle may occupy is mapped onto a
qubit associated with a unique location.

The dynamics of evolution in the QLG algorithm can be described in 
three scales, the microscopic, mesoscopic, and macroscopic scales.
At the microscopic scale, each particle has some probability of 
moving along the lattice. For example, the particle can move to the
right or left lattice site in a one-dimentional construction.  A 
simplified dynamics allows a particle to change its direction of
motion (via a {\it collision} with another particle) or keep moving 
at constant speed in its original direction of motion
({\it streaming}).  One can describe the dynamical behavior of 
particles at the mesosopic scale by determining their occupation probablities
on the lattice points; in one-dimention, only left-moving and 
right-moving probabilities are needed.  The mesoscopic dynamical
behavior of the system is modeled by a finite-difference form of a 
quantum Boltzmann equation.  Finally, to bridge to the macroscopic
scale, the occupation probabilities of the particles residing at each 
lattice site are summed together to determine the number density.
This number density quantity defined at each lattice node becomes a 
continuous field at the lattice resolution approaches infinity, which
is called the {\it continuum limit}.  Through a Chapman-Enskog 
purturbation procedure applied in the continuum limit, from the 
quantum
Boltzmann equation emerges an effective field theory that is 
parabolic in time and space and nonlinear in the number density
\cite{yepez-pl04}.

The QLG algorithm is initialized, in the NMR case, by encoding the 
particles' occupation probabilities as a spin-magnetization 
profile.
To handle the one-dimensional Burgers equation, it is sufficient to 
use two qubits (two spin-$\frac{1}{2}$ nuclei) per lattice site, where
each stores a single real valued occupation probability.
A room-temperature solution of isotopically-labeled chloroform ($^{13} 
CHCl_3$) was chosen for implementing the experiments,
where the hydrogen and the labeled carbon nucleus served as qubits 1 
and 2, respectively.  The difference of the gyro-magnetic
ratio of two spins generates widely spaced resonant frequencies that 
allows us to address each spin independently.
\begin{figure}[htbp]
\begin{center}
\epsfxsize=3.25in
         {\centerline{\epsffile{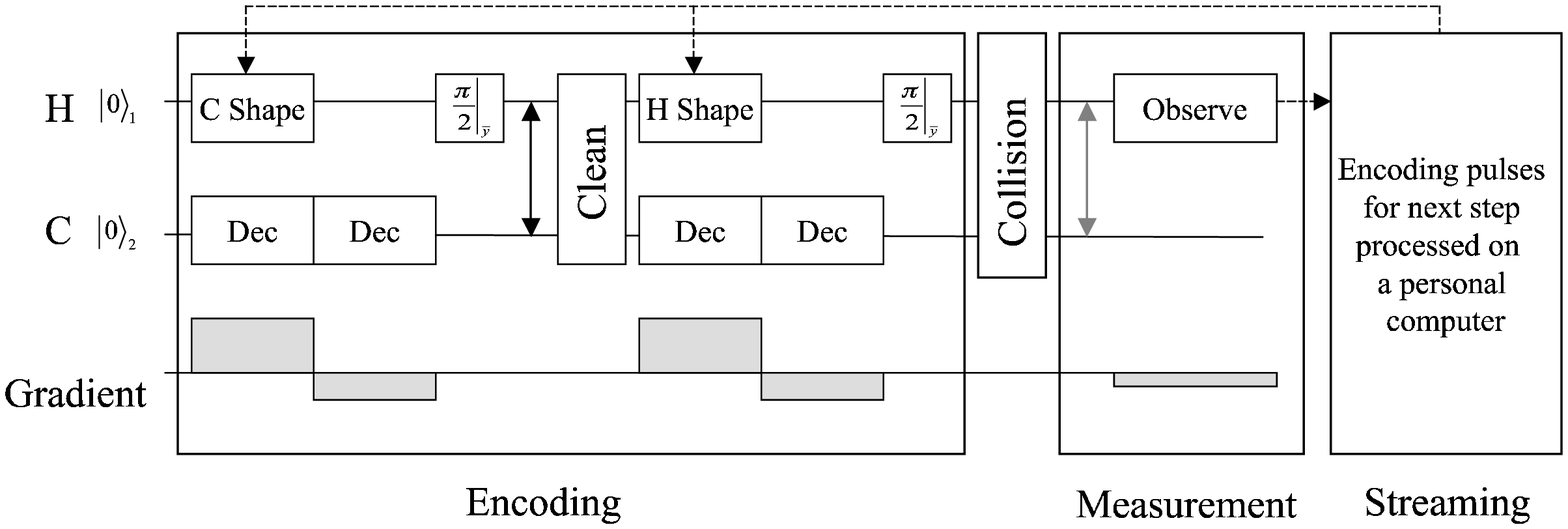}}}
         \caption{\footnotesize QLG algorithm implemented in four steps. Three
	  horizontal lines represent proton spin, carbon spin and 
fried gradients.  Both starting magnetizations
	  are encoded in proton channel first due to the high signal 
to noise ratio while decoupled in carbon channel to
	  prevent interfering of scalar coupling.  The collision 
operator is applied after the initialization.
	  Measurements are also taken in two steps in the proton 
channl followed by
	  data processing in a personal computer.}
         \label{Experiment Diagram}
\end{center}
\end{figure}
A lattice of QIPs are related to the ensemble sample by creating a 
correspondence between lattice sites and spatially
dependent positions in the sample.  A linear magnetic field gradient 
is used to generate distinct spatially-dependent
resonant frequencies that we can distinguish and modulate by a shaped 
RF pulse.  In this way, the magnetic
field gradient allows the entire spin ensemble to be sliced into a 
lattice of smaller, and individually addressable,
sub-ensembles.

The lattice initialization starts by transforming thermal equilibrium 
states into pseudo-pure states \cite{marco-ccc1043}.
The equilibrium state is highly mixed and the two nuclear spins have 
unequal magnetizations.  Thus, equalization of the
magnetizations is required prior to creating the pseudo-pure state. 
The dynamical evolution is caused by a collision operator
(a quantum operation), and measurement and streaming (classical 
operations) according to the QLG algorithmic paradigm.
The four main sections of the NMR implementation of QLG algorithm are 
graphically depicted in Figure~\ref{Experiment Diagram}.

First, each occupation probablility is mapped onto a lattice site as 
the expectation value of a number operator at a spacetime site
$n\Delta x$ at time $m\Delta t$.
As a result, the initial state of the $a^{th}$ qubit is
$\sqrt{f_{a}(n\Delta x,m\Delta t)} |1\rangle + \sqrt{1-f_{a}(n \Delta 
x,m \Delta t)}|0\rangle$.
The combined the wave function for a lattice site is a tensor product 
over the qubits,
\begin{eqnarray}
|\psi(n \Delta x,m \Delta t)\rangle & = & \sqrt{f_1f_2}|11\rangle +
  \sqrt{f_1(1-f_2)}|10\rangle \\
  \nonumber
   &+&\sqrt{(1-f_1)f_2}|01\rangle + \sqrt{(1-f_1)(1-f_2)}|00\rangle
\end{eqnarray}
In the basis of a two-qubit system, the number operators for the 
occupancy of qubits are defined in terms of the singleton qubit 
number operation $\hat n =\left(\begin{array}{cc}1&0\\  0 & 
0\end{array}\right)$ as follows: $\hat{n}_1 =  {\bf 1}\otimes \hat n$ 
and $\hat{n}_2 =  \hat n\otimes {\bf 1}$.
Therefore, the occupation probability is represented as follows:
\beqn{
f_{a}(n \Delta x, m \Delta t) = \langle\psi(n \Delta x,m \Delta 
t)|{\hat{n}_{a}}|\psi(n \Delta x,m \Delta t)\rangle.
}
\eeqn
The macroscopic scale dynamical quantity of the quantum lattice gas 
is the number density, $\rho$, defined as the sum of the
occupancy probablity. The equilibrium occupation probabilities that we use are
\beqn{f^{eq}_a = 
\frac{\rho}{2}+e_a\frac{5}{8}\left[1-\sqrt{1-\left(\frac{32\rho}{25}\right)\left(1-\frac{\rho}{2}\right)}\right],}
\eeqn
where $e_a$ is $\pm 1$ for different qubits.

The initial magnetization is specified by using a RF pulse shaped by 
the Fourier
transform of the desired magnetization (tranform of the initial number 
density profile).  While applying the shaped pulse, a carbon 
decoupling sequence is performed to
prevent the scalar coupling from interfering with the low power 
shaped pulses.  In addition, the $\frac{\pi}{2}$ pulse,
which rotates the information from the x-axis to the z-axis, is 
applied separately just after each initialization.
This is done to keep the valuable information along the longitudinal 
direction where it will not be affected
by the gradient and chemical shift.  The encoding of initial states 
on both spins is accomplished in two steps:
The initial carbon magnetication is recorded on the protons before 
being transferred to the carbons and followed by
the initialization of proton magnetication.  Furthermore, a short 
pulse sequence, called the {\it clean sequence}, is executed after the
first swap gate to erase the phase distortion that may be caused by 
the decoupling sequence.

Second,  the evolution of $f_{a}$ is governed by the combined action 
of the collision operator, measurement and
streaming.  The collision operator is applied to all the lattice 
sites independently, resulting in $|\psi^{'}(n \Delta 
x)\rangle=\hat{C} |\psi (n \Delta x)\rangle$, for all $n$.
The choice of the particular components of the unitary collision 
operator determines the form of the macroscopic effective field 
theory (a parabolic partial differential equation) and the value of its 
transport coefficients (coefficients of the dissipative terms).
A general representation of the collision operator for the Burgers 
equation is a block diagonal
matrix. This single quantum operator is chosen to be
\beqn{\hat C = \exp\left[-i\frac{\pi}{4.882} 
\left(\sigma_x^H\sigma_y^C-\sigma_y^H\sigma_x^C\right)\right]},
\eeqn
which has the following matrix representation:
\beqn{
\hat{C} = \left(
\begin{array}{cccc}

        1 & 0 & 0 & 0 \\
        0 & 0.8 & 0.6 & 0\\
        0 & -0.6 & 0.8 & 0\\
        0 & 0 & 0 & 1

\end{array} \right).}
\eeqn

The unitary operator $\hat C$ can be decomposed of a sequence of RF 
pulses and scalar coupling.
The product operators in the exponent commute with each other, 
resulting in $\hat C = \exp\left[-i\frac{\pi}{4.882} 
\sigma_x^H\sigma_y^C\right] \exp\left[-i \frac{\pi}{4.8828} 
\sigma_y^H\sigma_x^C\right]$.
  Both terms can be expanded as natural scalar Hamiltonian couplings 
sandwiched with the appropriate single rotations, resulting in
\begin{eqnarray}
\label{collision-operation-pulse-form}
  \hat C & = & e^{-i \frac{\pi}{4} (\sigma^H_y + \sigma^C_y)}
       e^{-i \frac{\pi}{4} \sigma_z^H \sigma_z^C }
       e^{-i \frac{\pi}{4.882} (\sigma_x^H - \sigma_x^C)} \\
       \nonumber
       & \times &e^{i \frac{\pi}{4} (\sigma_y^H + \sigma_y^C)}
       e^{-i \frac{\pi}{4} (\sigma_x^H + \sigma_x^C)}
       e^{-i \frac{\pi}{4} \sigma_z^H \sigma_z^C }  e^{i \frac{\pi}{4} 
(\sigma_x^H + \sigma_x^C)}.
\end{eqnarray}
The exponential terms of single spin rotations are implemented by 
$\pi/2$ and $\pi/4$ pulses. The exponents
of terms with $\sigma_z^H\sigma_z^C$ represent the natural internal 
Hamiltonian evolutions with time period $1/2J$, where
$J$ is $214Hz$.
Here, the evolution of the internal Hamiltonian is ignored while the 
RF pulse is applied. This approximation leads to a
systematic error that will accumulate during the course of the computation. 
In general, these errors are easy to avoid, but
since the purpose of the investigation was to explore the sensitivity 
to accumulated errors we did not correct it.
The collision operator follows the encoding (Step 2), and it is 
implemented without magnetic field gradients to ensure
that all of the sites in the sample undergo the same transformation.

Third, we measure the occupation probabilities.  This process erases 
all the superpositions and quantum entanglement that was created by 
the unitary collision operator in the second step.

The occupation numbers of each spin are obtained following the 
collision step by measuring the z-magnetization
according to the following equation
\beqn{ f_a (n,m) = \frac{1}{2}\left[1+\langle\psi(n,m)| \sigma_z^a 
|\psi(n,m)\rangle\right] }.
\eeqn
Since only $\sigma_x$ and $\sigma_y$ are observable in our NMR 
spectrometer, a $\pi/2$ pulse has been used to bring
the z-magnetization into the transverse plane. The measurements are 
done in two seperate experiments, where a SWAP gate
is applied to bring the magnetization from carbon channel to the 
proton channel. This SWAP operation is done because the
higher signal-to-noise ratio in the proton channel allows us to 
improve the accuracy of our implementation.  During the
``readout'' process (Step 3), a week magnetic field gradient is 
applied to distinguish different sites.  The observed
proton signals are digitized and Fourier transformed, allowing us to 
record the spatially-dependent spin magnetization profile.

Fourth, and last step of the QLG algorithm, we shift the $f_{a}$ obtained 
in the previous step to  its nearest neighbor using a short  {\sc 
Matlab} program.  This step requires only classical communication 
between neighboring sites.  The time is incremented after this step. 
Then, we loop back to step 1 and update the field of occupation 
probabilities over the lattice sites. In
this way, we can continue to iterate forward in time and make a 
time-history record of the occupation probabilities, which in turn 
gives us the temporal evolution of the number density field.
\begin{figure}[htbp]
\begin{center}
\epsfxsize=2.5in
        {\centerline{\epsffile{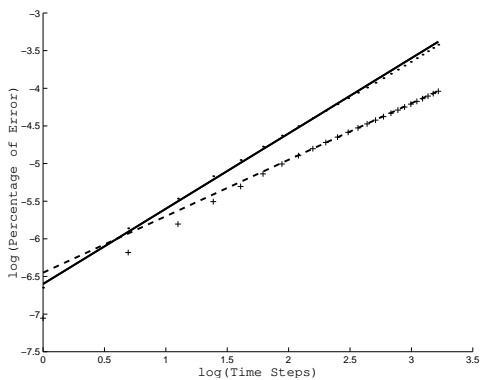}}}
         \caption{\footnotesize The growth of the systematic errors 
due to the collision operator in two NMR implementation.  The single 
collision operator data (dots) is fit (solid line)  with a line of 
slope $1$, which shows linear growth of the error.  The collision 
operator data  with modulated phases (pluses)  is the fit with a line 
of slope $3/4$ (dashed line).
	  The buildup of the systematic errors has been slowed down 
by proposed method.  However, the systematic errors
	  have not been totally converted into random errors.}
         \label{Error Analysis}
\end{center}
\end{figure}
In the implementation of the Burger equation, we observed deviations 
between the numerically predicted data points and
analytically predicted solutions.  These errors can be attributed to 
imperfections in the NMR implementation.  The major
error sources in the NMR implementation are known, so to explore the 
source and relative strength of these errors,
we have simulated the NMR experiments.  The major error source in 
this implementation is the collision operator, and
it is introduced by ignoring the scalar coupling between proton and 
carbon during the RF pulses.  When applying an
RF pulse on the proton qubit, the Hamiltonian in the rotating form is 
$H = 2 \pi J \sigma^H_z \sigma^C_z + \gamma_H B_1 \sigma^H_x$,  where 
$B_1$ is the strength of the RF pulse.  With the presence of the 
scalar coupling, a small portion of the proton
magnetization has been transfered to the carbon qubit.  Therefore, 
the applied propagator can be recast as $U = U_{\hbox{\tiny 
desired}}U_{\hbox{\tiny error}}$.

The error in the collision operator is a systematic error that builds 
up throughout the successive time steps.  Although
this is not the dominant error at the beginning of the 
implementation, it eventually dominates the first-order error due
to the Fourier approximation and becomes the dominant issue after just several time 
step interations.  Notice that while the reduction of the initial 
magnetization from the Fourier transform is systematic, since the 
magnetization profile is changing the errors are not precisely 
repeated.  In the collision operator, however, the errors are exactly 
the same from step to step.  In addition we expect that the radio 
frequency inhomogeneity leads to strongly correlated errors in the 
lattice encoding.  Hence, we have
proposed replacing a single collision operator with a set of 
collision operators that have similar fidelity but randomized
error terms.

Since the collision operator for the Burgers equation is a zero-order 
coherence term, the collision operator commutes with
the rotation operator.  Therefore, we apply a $90^{\circ}$ rotation 
operator to the collision operator at each step to mitigate error 
growth.  Consequently, a dramatic improvement is observed as shown
in Figure 2.  On a logarithmic plot, the simulation results fit a 
line with a slope of $3/4$.  If the
error terms in the collision operators were totally randomized and 
hence followed a Gaussian distribution, the best-fit
regression line should have had a slope of $1/2$.  The devitation 
between our simulation data and the ideal Gaussian case
indicates residual systematic error in the collision operator.  In a 
future study, we may use strongly mudulated pulses to randomize the 
error terms.
\begin{figure}[htbp]
\begin{center}
\epsfxsize=2.5in
{\epsffile{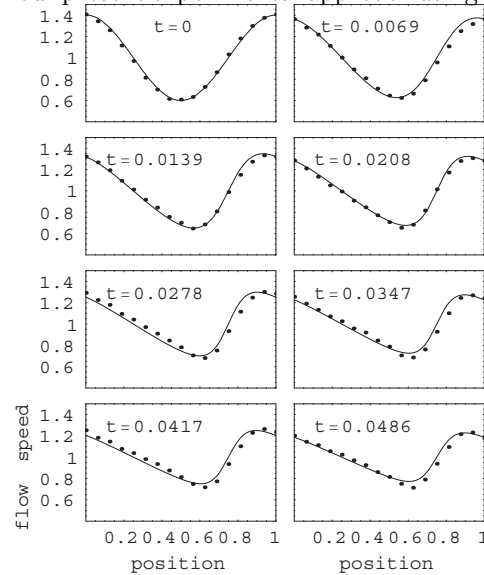}}
         \caption{\footnotesize The experimental data are plotted 
together with the analytical solutions for 8 time steps on a lattice
	  of 16 parallel two-qubit QIPs.  Viscosity: 
$\frac{1}{4}\frac{\Delta x^{2}}{\Delta t}$. Experimental NMR data 
(dots) versus analytical solution (curves).  Randomizing the error 
terms in the
	  collision operator has improved the experimental results 
dramatically. }
         \label{Experimental Data}
\end{center}
\end{figure}
The experimental number densities are over-plotted in 
Figure~\ref{Experimental Data} with the exact analytical solutions.
Eight  successive time steps of the quantum algorithm were 
implemented on 16 two-qubit sites.  An improvement of our
present experimental approach using collision operators with 
modulated phases is observed.  The agreement of the
data to the analytical solutions is encouraging and suggests that 
totally randomizing error terms in the collision operator may offer 
further improvement.

NMR quantum simulations has provided an alternative way to study the 
NMR spectroscopic implementations.  From the simulation,
we find the major error sources are due to imperfect control of the 
quantum spin system and the Fourier approximation
associated with setting its magnetization profile. Our proposed 
method for converting the systematic errors into random errors
is effective. The improvement we achieve relative to the previous 
experiment is encouraging, and it demonstrates the possibility
of using the same technique in future studies.
The closeness of the numerical data to the exact analytical results 
for the nonlinear Burgers equation further proves the
practicality of implementing the QLG algorithm using a spatial NMR 
technique.  In addition, although the limitation of the
Fourier approximation is not dominant, the problem of precisely 
initializing a lattice of QIPs still remains an open issue.

We thank M. Pravia, N. Boulant, H. Cho and Y. Liu for valuable 
discussion.  This work was supported by the Air Force
Office of Scientific Research, along with DARPA, ARDA and ARO.

\end{document}